%% file: main.tex
\documentclass[sigconf,9pt]{acmart}

\AtBeginDocument{%
  \providecommand\BibTeX{{%
    \normalfont B\kern-0.5em{\scshape i\kern-0.25em b}\kern-0.8em\TeX}}}
\usepackage{wrapfig}
\usepackage{color}
\usepackage{array}
\usepackage{float}
\usepackage{graphicx}
\usepackage{enumitem}
\newcolumntype{P}[1]{>{\centering\arraybackslash}p{#1}}
\usepackage{hyperref}
\usepackage[export]{adjustbox}
\usepackage{multirow}
\usepackage{hyperref}
\usepackage{subfigure}
\usepackage[export]{adjustbox}
\usepackage{multirow}
\usepackage{subfigure}
\usepackage{caption}

\copyrightyear{2022}
\acmYear{2022}
\setcopyright{rightsretained}
\acmConference[SIGCOMM '22 Demos and Posters]{ACM SIGCOMM 2022 Conference}{August 22--26, 2022}{Amsterdam, Netherlands}
\acmBooktitle{ACM SIGCOMM 2022 Conference (SIGCOMM '22 Demos and Posters), August 22--26, 2022, Amsterdam, Netherlands}\acmDOI{10.1145/3546037.3546063}
\acmISBN{978-1-4503-9434-5/22/08}

\begin{document}
\captionsetup[figure]{labelfont={bf},name={Fig.},labelsep = colon,font=footnotesize}

\title{Terra: Blockage Resilience in Outdoor mm-Wave Networks}

\author{Santosh Ganji, Jaewon Kim, P. R. Kumar}
\affiliation{\small{Texas A $\&$ M University}}

\setlength{\belowcaptionskip}{-14pt} 
\input{abstract}


\maketitle
\input{intro}    
\input{background}
\input{methodology}
\input{results}

\input{conclusion}

\bibliographystyle{ACM-Reference-Format}
\bibliography{bib}
\end{document}

%% file: abstract.tex
\begin{abstract}
We address the problem of pedestrian blockage in outdoor
mm-Wave networks, which can disrupt Line of Sight (LoS)
communication and result in an outage.
 This necessitates either reacquisition as a new user that can take up to 1.28 sec in 5G New Radio, disrupting high-performance applications, or
  handover to a different base station (BS), which however
 requires very costly dense deployments to ensure multiple base stations are always visible to a mobile outdoors.
 
 We have found that there typically exists a strong ground reflection from concrete and gravel surfaces,
 within 4-6 dB of received signal strength (RSS) of LoS paths. The mobile can switch to such a Non-Line Sight (NLoS) beam to sustain the
 link for use as a control channel during a blockage event. This allows the mobile to maintain time-synchronization with the base station, allowing it to revert to the LoS path when the temporary blockage disappears.
 
 We present a protocol, Terra, to quickly discover, cache, and employ ground reflections.
It can be used in most outdoor built environments since pedestrian blockages typically last only a few hundred milliseconds.
 \end{abstract}

%% file: intro.tex
\section{Introduction}

\label{sec:intro}
Millimeter-wave (mm-Wave) spectrum offers huge bandwidth making possible data rates of the order of gigabits per second that are necessary for applications such as Virtual reality, Augmented reality, and high definition online gaming. 
However, poor propagation characteristics pose hurdles. 
mm-Wave communication systems must use narrow directional beams to combat high propagation loss in the $28$ to $80$ GHz frequency range. Line of Sight (LoS) beams are therefore used to maximize signal strength.
The challenge we address is to sustain the links outdoors during pedestrian blockages which attenuate the link by 15 dB leading to outage \cite{unblock}.
A mobile that loses connectivity with the BS will need to be reacquired as a new user.
In 5G New Radio, the BS sweeps 
 broadcast information in $64$ different directions every $20$ ms, and so reconnecting as  a new user
takes up to 1.28 sec plus the time
 to complete initial access procedures.
 The entire process is also energy exhaustive and disrupts user experience of low latency applications.

A solution to preserve connectivity in indoor environments during  blockage is to communicate along NLoS paths reflected from walls and myriad surfaces with RSS typically 8 to 10 dB less than LoS. This is
sufficient to sustain control plane communication and time synchronization between a mobile and BS. Solutions and protocols have been developed \cite{unblock,Ish_two_beams} to provide mechanisms to discover NLoS beams in indoor environments and sustain links that can recover to their LoS beams when the blockage disappears.

Under the premise that NLoS paths are not available in outdoor environments, it is commonly suggested to either perform handover to a nearby BS \cite{Ish}, or to employ Coordinated Multipoint Transmission (CoMP) \cite{BS1,BS2}.
For the latter, at least three BSs must share high-resolution channel state information (CSI) \cite {BS1,BS2,comp}.
Both approaches require dense deployment of BSs with overlapping coverage. A study \cite{Ish} shows that a 200 $BS/km^2$ 5G New Radio BS density is necessary to meet the performance of low latency applications with handoff -- a capital intensive deployment.
 Moreover, since pedestrian blockage is an unpredictable event, coordinating BSs must share CSI for all the active users  via a side-channel or backhaul with end to end communication latency less than a symbol duration,
 which can be as little as 9 $\mu$s in 5G NewRadio, a challenging proposition.

  \captionsetup{singlelinecheck = false,  justification=raggedleft, font=footnotesize, labelsep=space}
\setlength{\columnsep}{-1.25in}%
 \begin{wrapfigure}[12]{r}{0.6\linewidth}
\begin{figure}[H]
   \raggedleft
    \includegraphics[width=.6\linewidth,keepaspectratio]{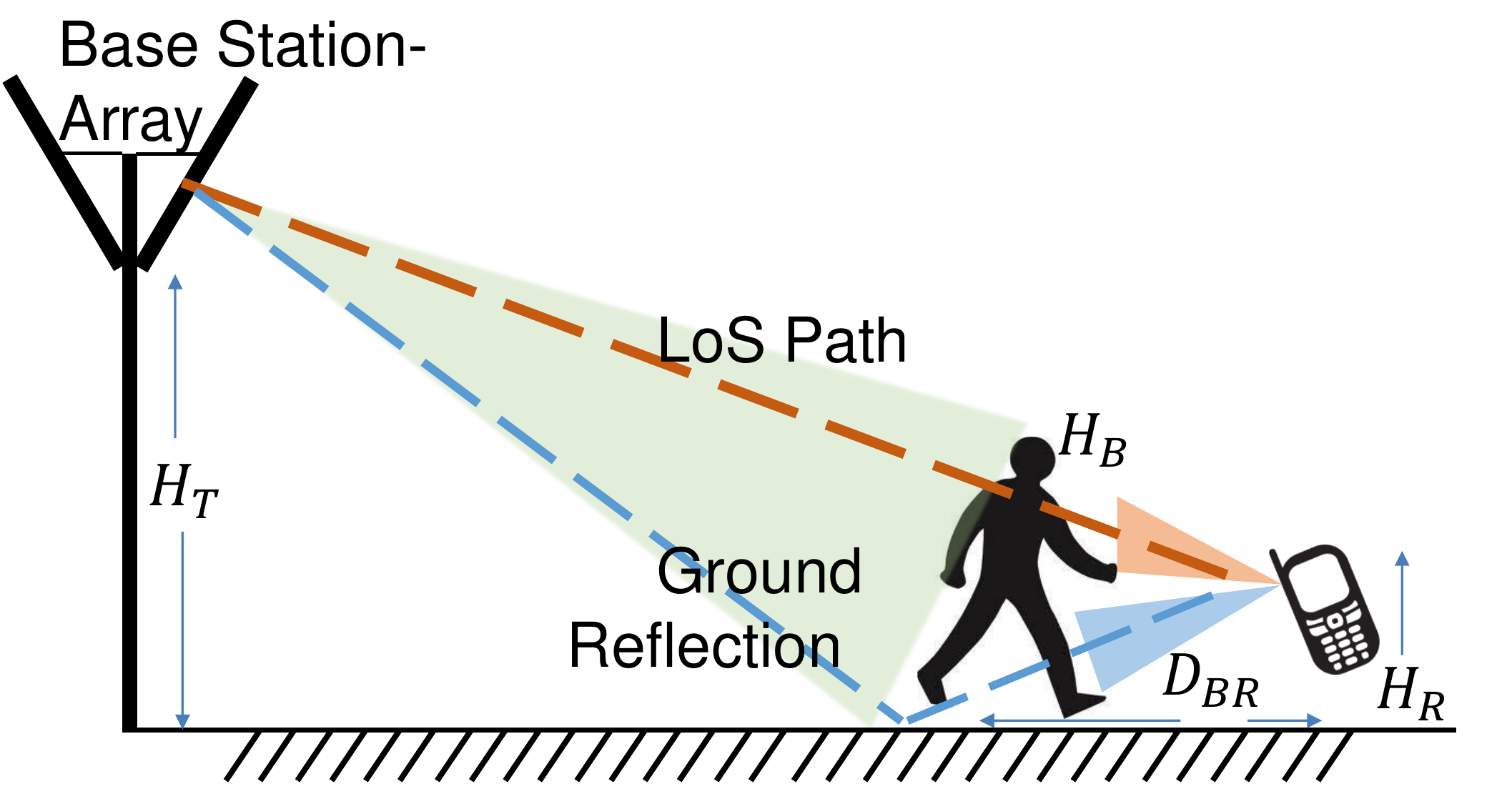}
    \caption{\begin{small}{Ground Reflection}\end{small}}
    \label{GR}
\end{figure} 

\end{wrapfigure}

  \captionsetup{singlelinecheck = false, format= hang, justification=centering, font=footnotesize, labelsep=space}

\begin{figure*}[t!]
\hspace{-1cm}

\begin{minipage}[b]{.37\textwidth}
\includegraphics[width=1.02\linewidth]{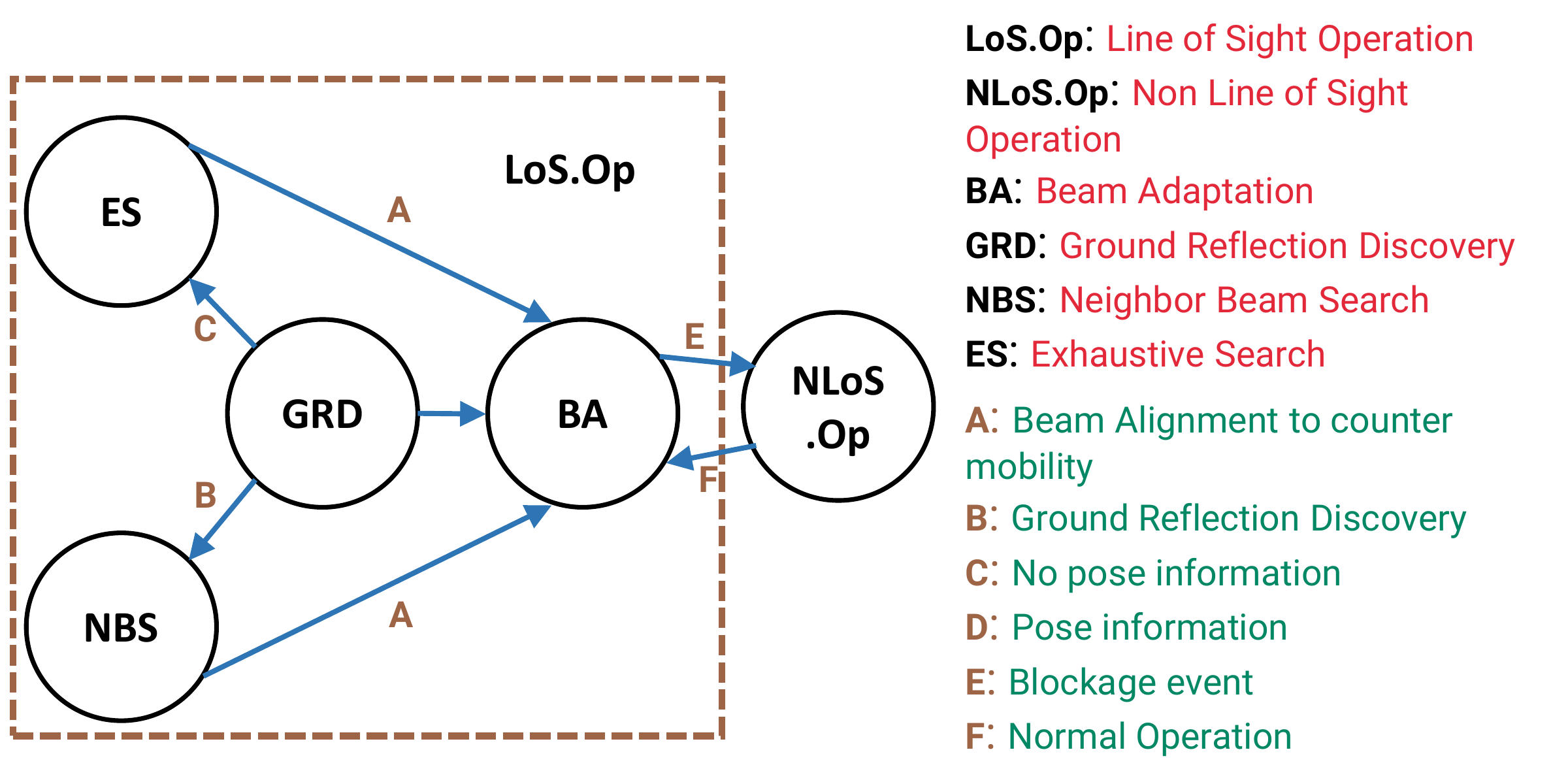}
\caption{\begin{small}State machine\end{small}} \label{terra}
\end{minipage}
\hspace{.5cm}
\begin{minipage}[b]{.21\textwidth}
\includegraphics[width=1.18\linewidth]{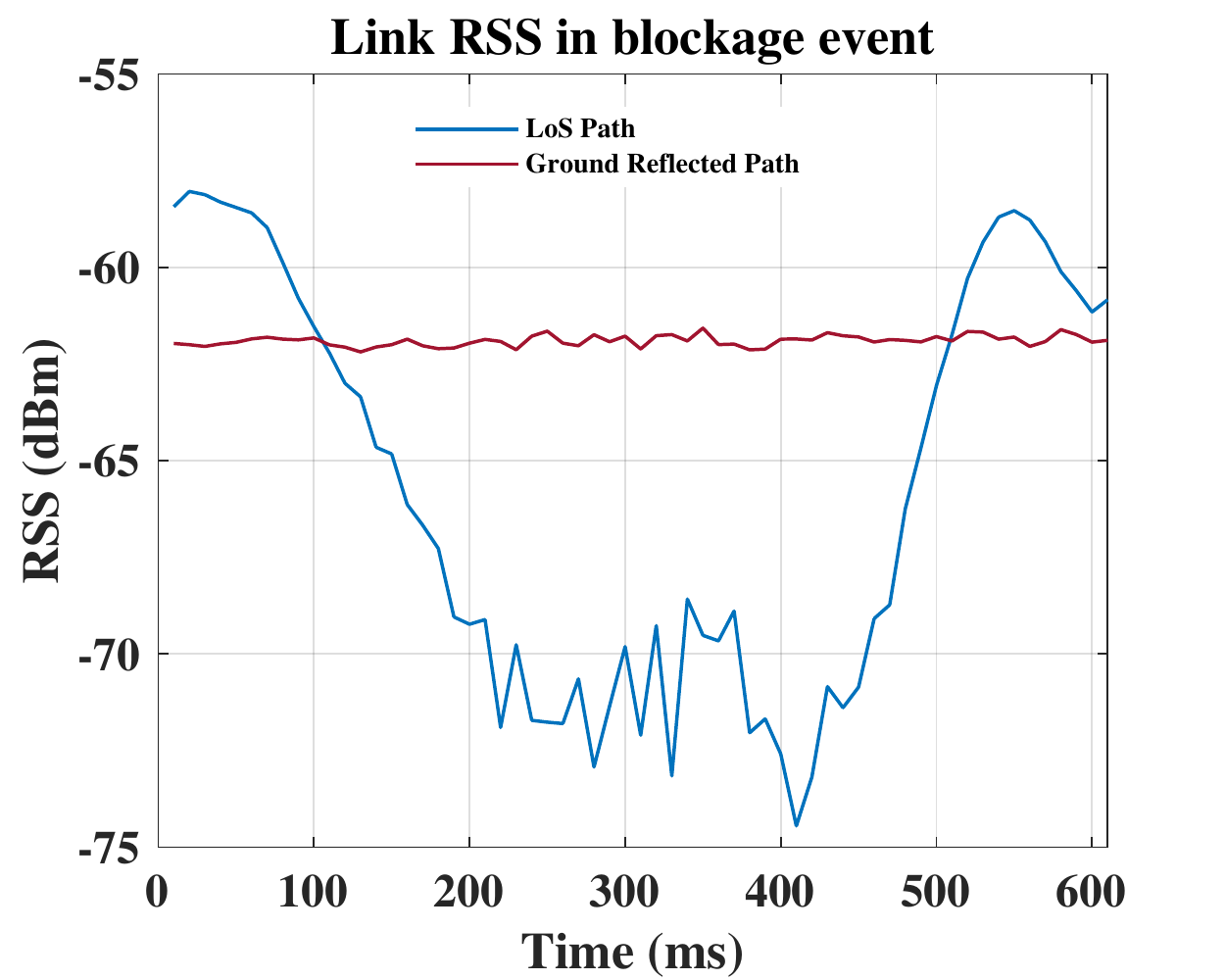}
\caption{\begin{small}{Blockage}\end{small}} \label{blockage}
\end{minipage}
\hspace{.5cm}
\begin{minipage}[b]{.21\textwidth}
\includegraphics[width=1.18\linewidth]{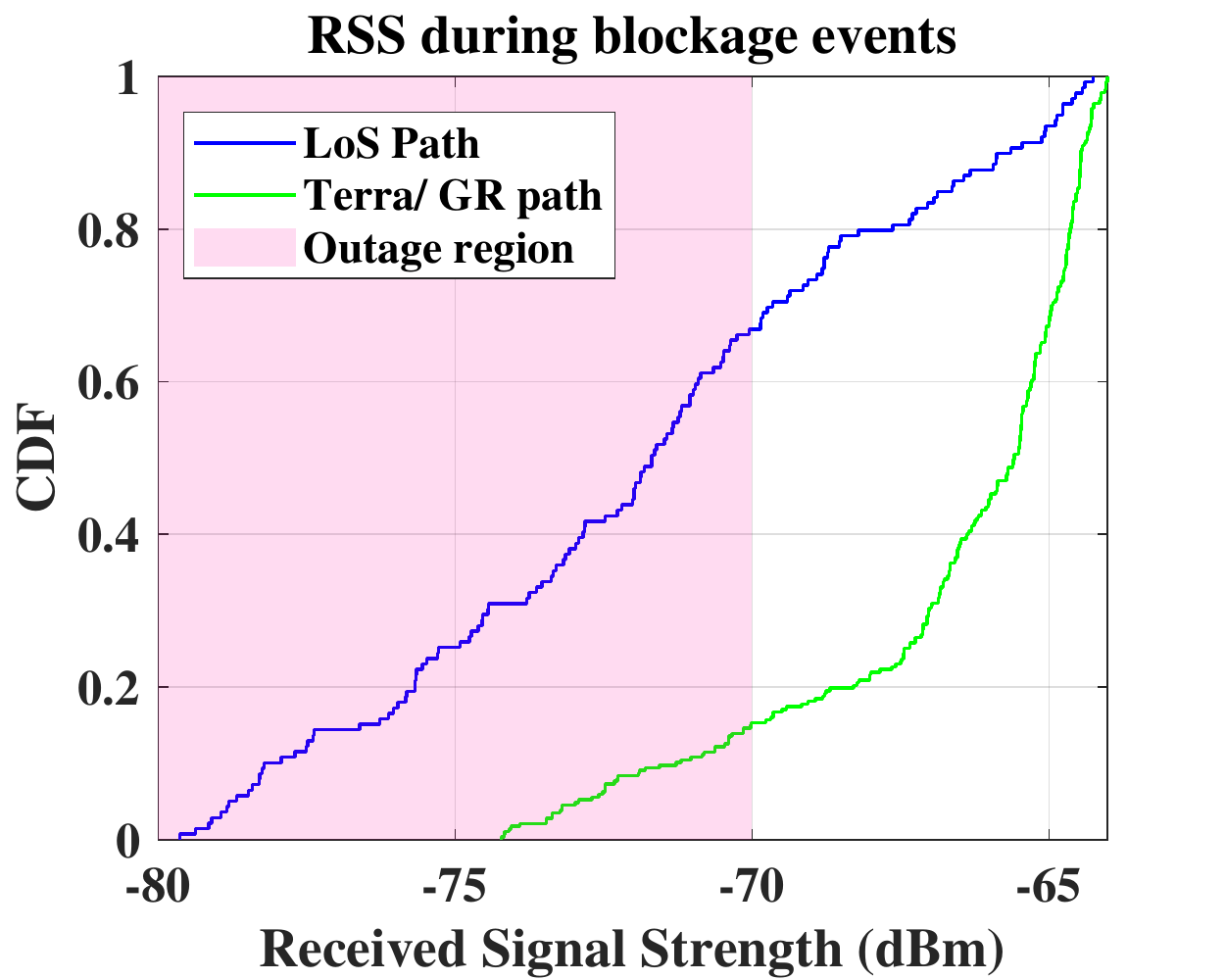}
\caption{\begin{small}Performance\end{small}} \label{performance}
\end{minipage}
\end{figure*}
We experimentally show that there is a simple alternative solution in many outdoor built environments.
Through an extensive measurement campaign 
in the ISM band at $60$ GHz, where oxygen absorption is highest, \cite{blockage_human},
we have found omnipresent ground reflections
from hard surfaces such as concrete or gravel with an RSS about 4-6 dB less than LoS. If not high date rate communication, these ground reflected paths can support lower rate bi-directional control channel communication that can sustain time synchronization between mobile and BS which allows rapid switching back to the LoS beam as soon as the blockage disappears. Such reflections appear not to have been considered previously for mm-Wave communication. 
We specifically design and demonstrate a protocol, TERRA, to discover such ground reflections, sustain time-synchronization, and revert to LoS transmission as soon as a temporary pedestrian blockage, which typically lasts a few hundred milliseconds, ends. 

%% file: background.tex
\section{Background and Related Work}
The presence of ground reflections has been noticed earlier, but apparently not thoroughly investigated for mm-Wave communications. Rajagopal et al \cite{NLOSmain1}
characterized penetration and reflection losses for various materials, and
fleetingly mention the presence of ground reflections from concrete at $28$ GHz,
which from the presented graphs appear to be about
$6-8$ dB below LoS RSS.
Another outdoor measurement campaign reported in \cite{Rappaport} specifically mentions NLoS paths from the
scatterers including ``lamppost, building, tree, or automobile", but not the ground itself, and reports existence of NLoS paths that have 10-50 dB greater loss than LoS.

%% file: methodology.tex
\section{Experimentation}
\label{Experimentation}
The measurement campaign studied the additional path loss of ground reflection from ceramic tiles, concrete, and gravel surfaces. 
We used National Instruments $60$ GHz software-defined radios (SDRs) \cite{mmwavesdr} with A 2 GHz bandwidth OFDM waveform. Each transceiver has FPGA cards for base band IQ processing, A to D converters, and a 12 element rectangular phased array generating 25 beams spanning a $120^\circ$ sector. The zenith and azimuth beamwidths of beams are approximately $60^\circ$ and $18^\circ$, respectively. 
Transmit power and directivity gain are 20 dBm and 17 dB.
We set up the transmitter antenna 2.5 m above ground and receiver antenna array 1 m from the concrete surface.
The distance between transmitter and receiver was varied from 5 to 25 m.
To receive the ground reflected signal, transmitter and receiver arrays are tilted downwards.

We studied scenarios
where 
a human subject is interposed between transmitter and receiver.
Let $H_T$ and $H_R$ denote the distances from ground level to transmitter array and receiver array, $D_{TR}$ the distance between transmitter and receiver, and
$H_{B}$ the height of a human blocker. When $H_T >H_B> H_R$, only blockers close to the receiver can occlude the link. 
In one case, where $H_R=1$m, $H_T=2$m,  $H_B$= 1.78m, and $D_{TR}=6$m, pedestrians within 3m from the receiver block LoS path. In comparison to LoS RSS, 
The median additional loss over LoS along the ground reflected path is 4.5 dB and 4.8 dB, for concrete and gravel surfaces. Downward angles to be applied at receiver to receive ground reflection remain close for all the locations where pedestrian occlusions are possible. 

%% file: results.tex
\section{The Terra Protocol}
\label{sec:eval}
Terra is a protocol to discover ground reflected paths quickly with minimal usage of limited network and modem resources, and switch to them upon blockage.
The protocol must always have a usable NLoS path since
Pedestrian blockages are unpredictable and suddenly attenuate RSS by 10s of dB. 



Fig. \ref{terra} presents the State Machine of Terra.
In the ``Beam Adaptation" (BA) state it continuously
adapts its LoS beam to keep the BS and mobile's beams aligned under mobility, for which an 
algorithm such as \cite{BeamSurfer,twcbeamsurfer} or Agilelink \cite{Agilelink} can be used. In the ``Ground Reflection Discovery" (GRD) state, it obtains an NLoS beam
by searching over the neighboring zenith beams, called ``Neighbor Beam Search" (NBS), if mobile pose is available, or else by ``Exhaustive Search" (ES) over possibly all zenith beams till one is found.
As expected from geometry, the azimuth angle of the ground reflected path remains the same as that of the earlier LoS beam. As the user moves, Terra continually cycles through
BA and ground reflection beam discovery, ensuring that it always has both an LoS beam and an NLoS beam to the BS.
Terra stores the NLoS beam in its memory for future use in the ``NLoS Operation" (NLoS.op) state. 
Terra
continues operation in ``Line of Sight Operation" (LoS.Op) state either till blockage happens or until beam adaptation becomes necessary to adapt to mobility. The video demonstration of Terra shows its operation during human blockage \cite{Terravideo}.


%% file: conclusion.tex
\section{Evaluation}
\label{sec:conclusion}

We implemented Terra on NI mmWave SDRs \cite{mmwavesdr}. 
Fig.~\ref{blockage} shows a blockage event where the RSS of the LoS dipping below the noise floor, i.e., -70 dBm, during the blockage caused outage event that lasts for about 200 ms.
Fig. \ref{blockage} shows the RSS of both the LoS as well as the ground reflected paths as a pedestrian walks across the link on a concrete surface.
Fig. \ref{performance} plots the CDFs obtained by employing Terra during 50 blockage events.
The colored region shows the
outage region. The experiments show that Terra employs beams that are
outside the outage region $84.5\%$ of time, and within 6 dB of LoS operation $60\%$ of the time.
Terra either finds ground reflected radiation in just two measurements when pose information is available, or else it searches all the available 25 beams in the testbed \cite{mmwavesdr} till successful.

Post the above experiments, we then implemented and evaluated Terra on off-the-shelf 60 GHz routers \cite{mikrotik}. 
Without Terra 
the routers experienced 80$\%$ packet error rate in 50 pedestrian blockage events. In contrast, Terra maintained high link RSS with a packet loss of just 5$\%$ during blockage.